\documentclass[conference]{IEEEtran}
\usepackage{amsmath,amssymb,epsf,cite,times,graphicx}

\usepackage{pst-plot, pstricks}
\usepackage{xcolor}
\usepackage{amsmath}
\usepackage{amssymb}
\usepackage{graphicx}
\usepackage{epsfig}
\usepackage{float} 

\newcommand{\np}{n+\frac{1}{2}}

\newcommand{\sinc}{\textnormal{sinc}}
\newcommand{\nt}{\nabla\times}

\newcommand{\cwpnt}{\cos\tau\omega^n_p}
\newcommand{\swpnt}{\sin\tau\omega^n_p}
\newcommand{\scwpnt}{\sinc\tau\omega^n_p}

\newrgbcolor{mygreen}{0 0.5 0}
\newrgbcolor{myblue}{0.2 0.2 0.7}
\newrgbcolor{myred}{1 0.2 0.2} 

\newcommand{\ddt}[1]{\frac{d #1}{dt}}
\newcommand{\ppt}[1]{\frac{\partial #1}{\partial t}}

\renewcommand{\~}{\widetilde }
\begin{document}

\title{Three-dimensional relativistic particle-in-cell hybrid code
based on an exponential integrator}

\author{
	\IEEEauthorblockN{T. T\"uckmantel, A. Pukhov}
	\IEEEauthorblockA{ Institut f{\"u}r Theoretische Physik I \\
	Heinrich-Heine Universit{\"a}t D{\"u}sseldorf \\ 40225, D{\"u}sseldorf, Germany \\
	%Email: http://www.michaelshell.org/contact.html}
	}
	\and
	\IEEEauthorblockN{J. Liljo, M. Hochbruck}
	\IEEEauthorblockA{Mathematisches Institut \\ Heinrich-Heine Universit{\"a}t D{\"u}sseldorf \\
	40225, D{\"u}sseldorf, Germany }
}
%\author{T.~T\"uckmantel$^1$, J.~Liljo$^2$, M.~Hochbruck$^2$, A.~Pukhov$^1$}
%\affiliation{
%$^1$Institut f{\"u}r Theoretische Physik I, 
%Heinrich-Heine Universit{\"a}t, 40225, D{\"u}sseldorf, Germany
%$^2$Mathematisches Institut, Heinrich-Heine Universit{\"a}t, 
%}
\maketitle

\begin{abstract}
In this paper we present a new three dimensional (3D) full electromagnetic
relativistic hybrid plasma code  H-VLPL (hybrid virtual laser plasma
laboratory). The full kinetic particle-in-cell 
(PIC) method is used to simulate low density hot plasmas while
the hydrodynamic model applies to the high density cold background plasma.
To simulate the linear electromagnetic response of the high density
plasma, we use a newly developed form of an exponential integrator
method. It allows us  to simulate plasmas of arbitrary
densities using large time steps.
The model reproduces the  plasma dispersion and gives correct spatial
scales like the plasma skin depth even for large grid cell sizes. We
test the hybrid model validity by 
applying it to some physical examples. 
\end{abstract}
%\begin{keyword}
%\keywords{Plasma simulation; Object-oriented programming;
%Particle-in-Cell; hydrodynamic description; Hybrid simulation; Laser
%plasma interactions}
%\end{keyword}

%\pacs{52.65.-y, 52.65.Rr; 52.65.Kj, 52.65.Ww, 52.65.Kd} 

\section{Introduction}

It is well accepted now that the particle-in-Cell (PIC) codes provide
the most detailed description of plasmas and are the key computational
tools in the study of relativistic laser-plasma
interactions \cite{birdsal,hockney}. Large full 3D parallel
electromagnetic simulation codes like VLPL \cite{vlpl}, OSIRIS
\cite{osiris}, VORPAL \cite{vorpal}, OOPIC \cite{oopic}, and others
contributed remarkably in our understanding of the complex
laser-plasma physics. Because these codes provide the most detailed
plasma description, they are computationally expensive. As a result
one continues to look for new 
algorithms and simulation techniques to cope with
challenges of the laser-plasma physics.

One of the reasons why the classical explicit PIC
methods are computationally extremely expensive is that they have to
resolve the fundamental plasma frequency $\omega _p=\sqrt{4\pi n_e e^2/m_e}$,
which is the frequency of the plasma electrostatic oscillations. 
Therefore, they are efficient only when applied to low density
plasmas.  

At the same time, there is a number of important applications where
lasers interact with high density plasmas, e.g., the studies of
electron propagation through solid density targets and the resulting
target normal sheath acceleration (TNSA) 
\cite{solid_experiments}. The solid state density plasma
densities are in the range of $100-1000~n_c$, where 

\begin{equation}
n_c = m\omega^2/4\pi e^2 \label{eq:n_c}
\end{equation}
\noindent is the critical 
plasma density. Here, $m$ is the electron mass, $-e$ is its charge,
and $\omega$ is the laser frequency. For the 1~$\mu$m wavelength laser
the critical electron density is $10^{21}~$1/cc. Other important applications
include the Fast Ignition (FI) physics in the Inertial Confinement
Fusion (ICF) studies \cite{FI}. The FI plasma has a density of the 1000 times
compressed solid hydrogen, i.e., of the order of $10^5~n_c$. Hence,
the applicability of the classical PIC codes in this density range 
is facing severe difficulties. In this situation, one is forced to look for
a more efficient numerical method to challenge those ultra-high
densities. One of the possibilities is to include hydrodynamic
description of the high density plasma in the fully kinetic PIC code.

In last years PIC-hydrodynamic hybrid techniques have 
emerged as an efficient solution to large scale ultra high-density 
plasma simulations, e.g., FI physics, solid state 
density plasma interactions, high charge, high energy ion generations 
etc~\cite{Mason,Davies,Gremillet}. Most of these codes work in the
Darwin approximation and thus exclude the electromagnetic wave
propagation completely. They also exclude electrostatic waves keeping
the collisional magnetohydrodynamics (MHD) only. Further, an implicit 
electrostatic particle-fluid hybrid plasma code has been developed by Rambo and 
Denavit~\cite{rambo91}, which has been used to study interpenetration 
and ion separation in colliding plasmas~\cite{rambo94}. There is also the
implicit electromagnetic PIC code LSP 
\cite{LSP}. This code uses an implicit global scheme which overcomes
such restrictions of the time-step. The LSP code also employs a field
solver based on an unconditionally Courant-stable
algorithm\cite{Zheng} for electromagnetic calculations.  

Recently, we have presented a 1D version of the code  Hybrid Virtual
Laser Plasma Laboratory (H-VLPL) \cite{hvlpl} that unites a
hydrodynamic model for overdense plasmas and the full kinetic
description of hot low-density electrons and ions. In this code, the linear
plasma response was simulated using an implicit scheme. 
% That scheme simply cuts out the high frequencies down to those 
% corresponding to the time step. 
The implementation involved the solution of linear systems, which 
have been done in a very efficient way using the Schur complement. 

Unfortunately, the efficiency of the implicit H-VLPL code
drops significantly if we extend the code from 1D to 3D. Therefore, 
we introduce a new 3D version of the code that is based on a
different approach. Instead of using an implicit method, we employ a special
variant of an exponential integrator \cite{HocO10} to model the high frequency
plasma response. Exponential integrators are methods which make use of
matrix functions related to the matrix exponential of the
Jacobian of the differential equation. Here we consider a modification of the
mollified impulse method \cite{GarSS98}, which has been proposed for
molecular dynamics simulations.

The mollified impulse method is motivated from a splitting approach.
Variants of splitting methods are widely used for problems acting
on different time scales, see \cite{HaiLW06}. For our hybrid model it turns
out that due to the high density of the plasma, the highest frequencies stem
from a multplication operator, which acts only locally on each grid point.
Frequencies arising from the Maxwellian part are much lower and can be
handled explicitly as in the PIC code. This allows to
implement the new mollified impulse method by evaluating matrix functions
of diagonal matrices only. Obviously, this is much more efficient than
the solution of linear systems resulting from a 3D discretization.

% Here we split the problem
% into a Maxwellian part and a harmonic oscillator. In this case
% one can solve the resulting systems exactly which is, especially in 3D, a big
% advantage compared to solving a linear system as one would have to in an
% implicit scheme. The solution of the Maxwellian part is linear and the solution
% of the harmonic oscillator is given by the variation-of-constants formula. By
% this means we get an explicit scheme, where no solution of a linear system is
% required. In the variation-of-constants formula one has to compute the
% exponential of a matrix containing the high plasma density, but since that
% matrix is diagonal this does not pose a challenge. 

To illustrate the performance of the new method, we apply it to a few test physics
examples. We check the correct dispersion of electromagnetic waves in
the hybrid plasma and also compare the numerical skin length with the
analytical expressions. As a more complicated test, we apply the new
3D code to the laser-solid interaction and acceleration of protons in
the TNSA regime. Additionally, we perform a feasibility study of simulations 
of the Weibel instability, which occurs in the FI scenario.

The paper is organized as follows. First, we describe the full hybrid
method in Section \ref{sec:math}. Then, we briefly explain the numerical
scheme in Section \ref{sec:math_alg} and finally, we test the new
code H-VLPL on some well-known physical examples in
Section \ref{sec:tests}.

\section{Hybrid model}\label{sec:math}

In this section we define the physical model we use to simulate the
plasma. We use the full kinetic description with the usual PIC
macroparticles for low density hot electrons and ions. The high density cold
background plasma is then described hydrodynamically. Since the
spatial locations of the kinetic particles and the hydrodynamic parts
may overlap, we have to add currents generated by all species in the
same grid cells. 

The equations for the fields and particle momenta read
\begin{subequations}\label{eq:maxwell}
\begin{align}
\ppt{\textbf{E}} &=c\nabla\times \textbf{B}-4\pi\sum_\ell \textbf{j}_\ell,
\qquad\ell=e,i,h
\label{eq:maxwella}\\ 
\ppt{\textbf{B}} &=-c\nabla\times \textbf{E} \label{eq:maxwellb}\\
\ddt{\textbf{p}_h} &=q_e\textbf{E} \label{eq:maxwellc}\\
\ddt{\textbf{p}_\ell} &= q_\ell (\textbf{E}+\frac{\textbf{v}_\ell}{c} \times
\textbf{B}), \qquad\ell=e,i \label{eq:maxwelld} 
\end{align}
\end{subequations}
where 
\begin{align*}
 \textbf{j}_\ell=q_\ell n_\ell \textbf{v}_\ell, \qquad \textbf{p}_\ell=m_\ell
\gamma_\ell \textbf{v}_\ell, \qquad
  \gamma_\ell = \sqrt{1+\frac{\textbf{p}_\ell^2}{(m_\ell c)^2}}
\end{align*}
The index $\ell=e,i,h$ denotes electrons, ions, and
hybrid particles, respectively. $\textbf{E}$ and $\textbf{B}$ denote the electric
and magnetic field vectors, $\textbf{j}$ denotes the current density,
$\textbf{p}$ is the momentum and $n$ the number density of particles.

In the momentum equation (\ref{eq:maxwellc}) we have
neglected the nonlinear part of the Lorentz force ${\bf v \times
B}/c$, because we assume that the velocity of the electron part in the
cold background plasma is small, $v \ll c$. This assumption, however, limits the
cold plasma response to the linear one.

\section{Numerical algorithm}\label{sec:math_alg}

For simplicity, we rewrite the equations in dimensionless variables,
$\~t =\omega_0 t$ and $\~x= k_0 x$, where $\omega_0$ denotes the laser
frequency and $k_0=\omega_0/c$. The new variables are then
\begin{align*}
  {\~E} &= \frac{ eE}{m_ec\omega_0}, \quad
  \~B = \frac{ eB}{m_ec\omega_0}, \quad
  \~p_h = \frac{ p_h}{m_e c}, \quad
  \~p_\ell = \frac{ p_\ell}{m_\ell c}, %\\ \rm{and} \\
\end{align*}
and
\begin{align*}  
  \~j_\ell &= \frac{ j_\ell}{j_c}, \quad j_c = e n_c c, \quad
  \~n = \frac{n}{n_c}, \quad \~{q_\ell} = \frac{q_\ell m_e}{em_\ell}, \quad
  \~v_\ell = \frac{v_\ell}{c},
%, \quad \rho_c = \frac{m_e \omega_0^2}{4\pi q_e^2}.
\end{align*}
where $\ell=e,i$.

In the following, we omit the tildes, neglect the Lorentz force and consider hybrid
particles only. We can then write $\textbf{p}=\textbf{p}_h$. Using
these simplification, Eq.~\eqref{eq:maxwell} reads
\begin{subequations}\label{eq:maxwelldl}
\begin{align}
\ppt{\textbf{E}} &= \nabla\times\textbf{B}+\omega_p^2\textbf{p} \label{eq:maxwelldla}\\ 
\ppt{\textbf{B}} &=-\nabla\times\textbf{E} 		   \label{eq:maxwelldlb}\\
\ddt{\textbf{p}} &= -\textbf{E} \label{eq:maxwelldlc},
\end{align}
\end{subequations}
where $\omega_p^2 = \frac{n_h}{\gamma_h}$.

The problem is considered in three space dimensions. We solve the equations on a 
staggered grid and approximate the spatial derivatives with centered finite differences
using the Yee scheme \cite{Yee}.

\begin{figure}[H]
 \begin{center}
 \includegraphics[width=1.0\columnwidth,keepaspectratio]{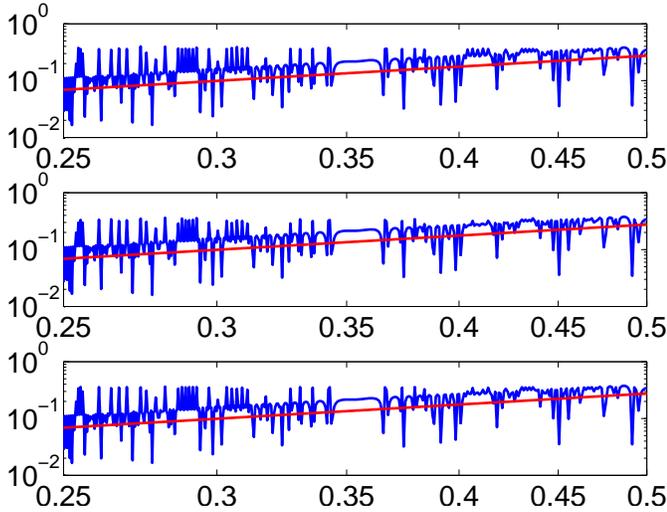}
\end{center}
\caption{Error in $E_y$, $B_z$ and $p_y$ plotted over the step size (blue) for
a straightforward integration of Eq.~\eqref{eq:split}. The red line shows 
the expected order two.}
\label{fig:nofilter}
\end{figure}

For the time discretization we will use the following splitting of the vector fields
\begin{align}
\begin{bmatrix} \dot{\textbf{p}} \\  \dot{\textbf{E}} \\ \dot{\textbf{B}}
\end{bmatrix} &=
\begin{bmatrix} -\textbf{E} \\ \nt \textbf{B} + \omega_p^2 \textbf{p} \\
   -\nt \textbf{E} \end{bmatrix}  \nonumber\\
  &= 
\begin{bmatrix} 0 \\ 0 \\ -\nt \textbf{E}
\end{bmatrix}+\begin{bmatrix} 0 \\ \nt \textbf{B} \\ 0 \end{bmatrix} +
\begin{bmatrix} -\textbf{E} \\ \omega_p^2 \textbf{p} \\ 0
\end{bmatrix} \label{eq:split}
\end{align}
If $\omega_p$ is constant over a time step, the exact solution 
of each of the three differential equations can be computed 
very efficiently, in particular without solving
any linear system. A symmetric splitting yields the following scheme
\begin{subequations}\label{eq:fescheme0}
\begin{align}
 B^{\np}  &= B^n - \frac{\tau}{2}\nt E^n 	  \label{eq:feschemea0}\\
(E^+)^n   &= E^n + \frac{\tau}{2}\nt B^{\np}    
\label{eq:feschemeb0}\\
\begin{bmatrix} p^{n+1}\\(E^-)^{n+1}\end{bmatrix} &= \begin{bmatrix} \cwpnt &
\tau\scwpnt\\-\omega_p^n\swpnt & \cwpnt\end{bmatrix}\begin{bmatrix}
p^n\\(E^+)^n\end{bmatrix}\label{eq:feschemec0}\\
E^{n+1} &= (E^-)^{n+1} + \frac{\tau}{2}\nt
B^{\np}\label{eq:feschemed-}\\
B^{n+1} &= B^{\np} - \frac{\tau}{2}\nt E^{n+1}.
\label{eq:feschemee0}
\end{align}
\end{subequations}
Although  this splitting method is of classical order two (since it
is a symmetric scheme), 
it suffers from resonances, which arise in $E$, $B$, and $p$ if the
density becomes large. In
fact, the errors of this
scheme are of order zero for certain time steps. 
We illustrate this effect by simulating a 1D plane wave. 
The incoming laser pulse is modeled via inhomogeneous, time dependent
Dirichlet boundary conditions and zero as initial values. A 
spatial grid size of $0.5$ for $x\in[0,200]$ is used. The hybrid density is
set to $n_h=10^8n_c$ and the system is integrated over the
time interval $[0,200]$. The blue curves in Fig.~\ref{fig:nofilter} 
shows the errors in $E_y$, $B_z$ and $p_y$ of the standard splitting method
\eqref{eq:fescheme0} as a function of the time step size, while the
red line corresponds to a second-order error behavior. To improve the
presentation, we only show the interval $[0.25,0.5]$ for the time steps, 
but we would like to emphasize that the same effects have been obtained
for much smaller time steps as well.

Similar resonance effects have also been observed for multiple time stepping schemes in
molecular dynamics simulations \cite{BieS93} and for numerical methods for solving
second-order differential equations \cite{GarSS98,HocL99,GriH06}. 
Motivated by these papers, we suggest to apply filter functions and
averaging operators to the Maxwellian part and modify the standard splitting
method \eqref{eq:fescheme0} in the following way
\begin{subequations}\label{eq:fescheme}
\begin{align}
 B^{\np}  &= B^n - \frac{\tau}{2}\nt\phi(\tau\omega_p^n) E^n 	  \label{eq:feschemea}\\
(E^+)^n   &= E^n + \frac{\tau}{2}\psi(\tau\omega_p^n)\nt B^{\np}    
\label{eq:feschemeb}\\
\begin{bmatrix} p^{n+1}\\(E^-)^{n+1}\end{bmatrix} &= \begin{bmatrix} \cwpnt & \tau\scwpnt\\-\omega_p^n\swpnt & \cwpnt\end{bmatrix}\begin{bmatrix} p^n\\(E^+)^n\end{bmatrix}\label{eq:feschemec}\\
E^{n+1} &= (E^-)^{n+1} + \frac{\tau}{2}\psi(\tau\omega_p^n)\nt
B^{\np}\label{eq:feschemed}\\
B^{n+1} &= B^{\np} - \frac{\tau}{2}\nt\phi(\tau\omega_p^n) E^{n+1}  
\label{eq:feschemee}.
\end{align}
\end{subequations}

\begin{figure}[H]
 \begin{center}
 \includegraphics[width=1.0\columnwidth,keepaspectratio]{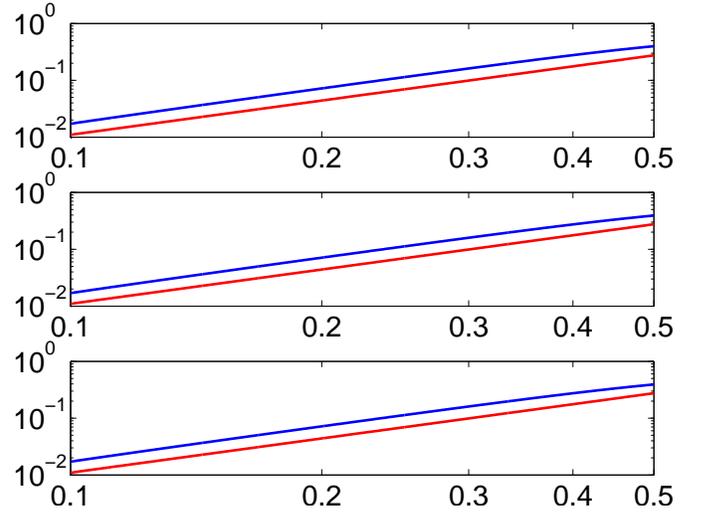}
\end{center}
\caption{Error in $E_y$, $B_z$ and $p_y$ plotted over the step size (blue) with
$\phi(x)=\psi(x)=\sinc~\frac{x}{2}$. The red line indicates the expected order
two.}
\label{fig:2filter}
\end{figure}

Fig.~\ref{fig:2filter} shows the same numerical test as for Fig.~\ref{fig:nofilter} with
filter functions $\phi(x)=\psi(x)=\sinc({x}/{2})$, where $\sinc(x):=\sin(x)/x$ and time
steps $\tau \in[0.1,0.5]$. 
The resonances have been eliminated completely and order two is achieved for arbitrarily
large densities. The theoretical properties of the numerical method including a detailed
error analysis is currently investigated and will be reported elsewhere.

\section{Benchmark with physical processes}\label{sec:tests}

The numerical integrator which was described in the previous sections has been implemented into the VLPL code as a three-dimensional, parallelized version, and is now
operational. In order to examine its accuracy and reliability, we have benchmarked it with a variety of physical processes. \\
First, we check if it correctly models laser propagation through
linearly dispersive plasma as well as reflection from overdense
plasma. Second, we verify the conservation of the total energy of the
system by the hybrid algorithm. Third, our code is applied to the
Target Normal Sheath Acceleration (TNSA) process, which would have
been very difficult to treat just using PIC means since it uses
materials of solid state density. We check if our hybrid integrator
correctly describes the exponential decay of a wave in overdense
plasma. Finally, we show its applicability to study the
Weibel instability. \\ 

\subsection{Reflection of an incident pulse}
As the simplest test one can imagine, we will show that our integrator
accurately models refraction in underdense plasma and reflection from
overdense plasma. 
First, we set up a plasma slab of $0.85 n_c$ density
(\ref{eq:n_c}) 
and send a $26 fs$ Gaussian laser pulse through it.
As the pulse hits the surface of the purely hybrid plasma, a part of the wave
is transmitted while a significant reflection also occurs.  
\begin{figure}[H]
	\begin{minipage}{0.48\columnwidth}
		\begin{center}
			\includegraphics[width=1.0\columnwidth]{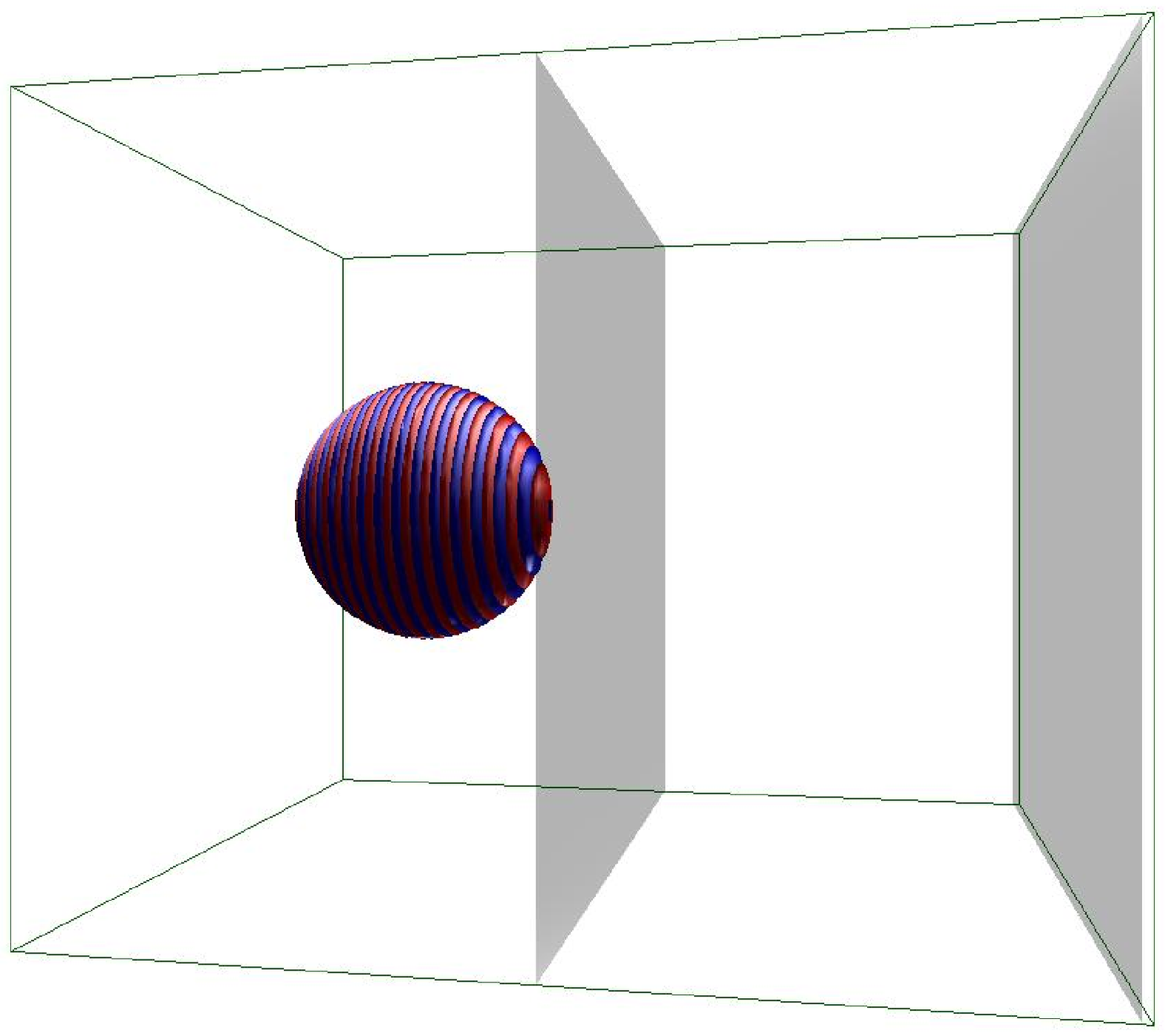}
		\end{center}
	\end{minipage}%
	\begin{minipage}{0.04\columnwidth}
		\hfill
	\end{minipage}%
	\begin{minipage}{0.48\columnwidth}
		\begin{center}
			\includegraphics[width=1.0\columnwidth]{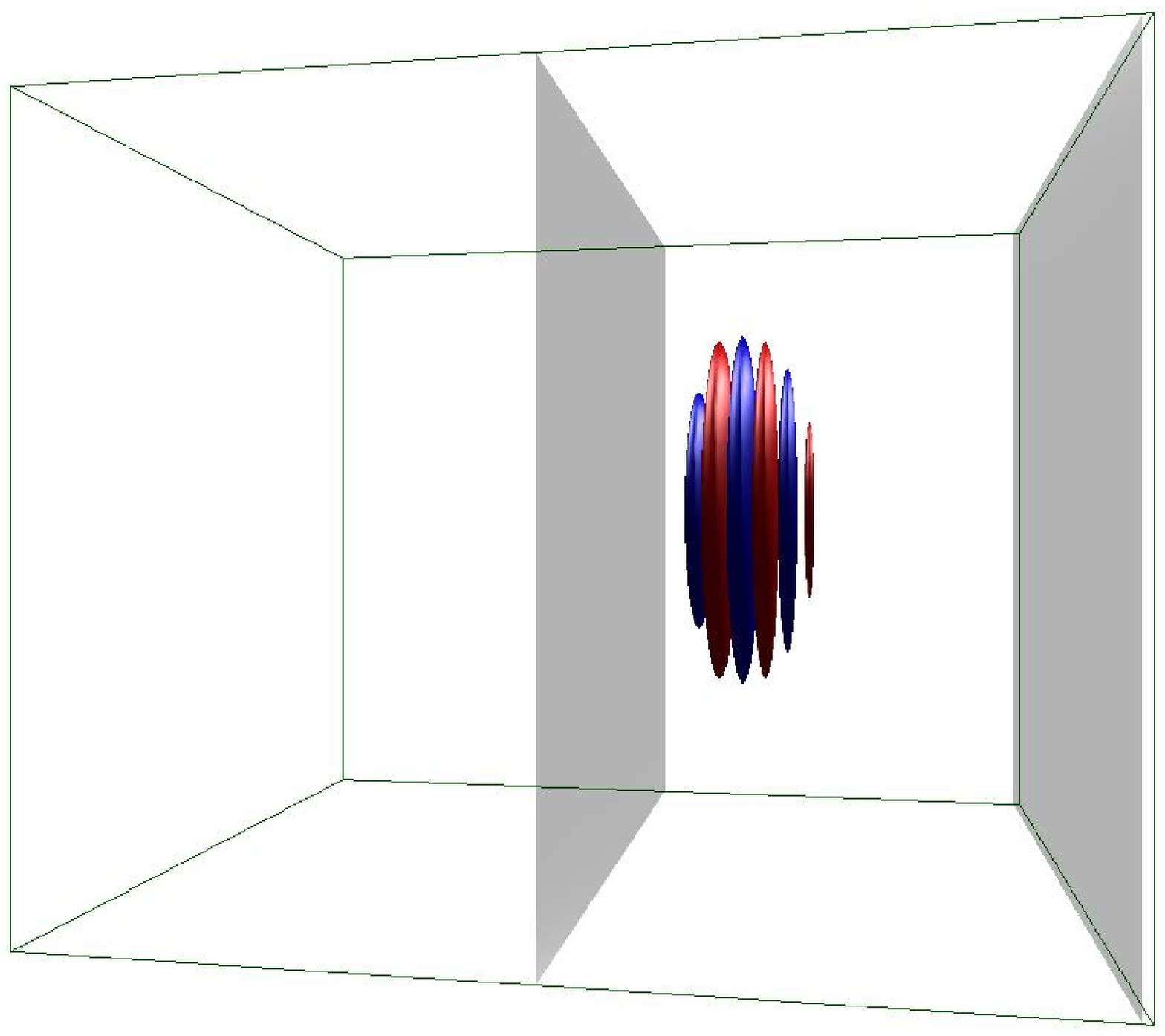}
		\end{center}
	\end{minipage}%
	\caption{Snapshots of the simulation setup taken with an interactive VR visualization software, which is a part of H-VLPL and currently under development. The left picture 
		shows the laser pulse (isosurfaces of fixed positive and negative electric field amplitudes) as it enters the hybrid plasma. The right picture demonstrates the dispersive
		effect. }
	\label{disp}
\end{figure}

%\begin{figure}[H]
%		\begin{center}
%			\includegraphics[width=1.0\columnwidth]{frame_disp000.ps}
%			\includegraphics[width=1.0\columnwidth]{frame_disp001.ps}
%			\caption{Snapshots of the simulation setup taken with an interactive VR visualization software, which is a part of H-VLPL and currently under development. The left picture 
%				shows the laser pulse (isosurfaces of fixed positive and negative electric field amplitudes) as it enters the hybrid plasma. The right picture demonstrates the dispersive
%				effect. }
%			\label{disp}
%		\end{center}
%\end{figure}
On the other hand, when the experiment was modified by setting the density to $1.2 n_c$, we observe a reflection of the entire electromagnetic wave by the plasma. \\
We point out that these simulations have been performed using just the fluid part of our combined code without any PIC macroparticles. Still, the effect has been described
correctly.

\subsection{Energy conservation}
\begin{figure}[H]
	\begin{center}
		\includegraphics[width=1.0\columnwidth]{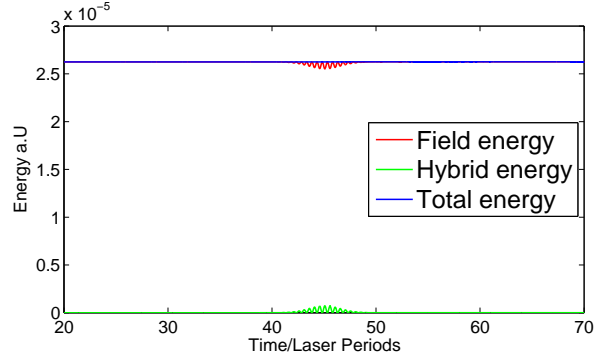}
		\caption{Plot of the total energy, the energy of the electromagnetic field and of the hybrid plasma versus time.}
		\label{energy}
	\end{center}
\end{figure}

An important property we require from the new integrator is the conservation of the total energy of the system, comprising PIC macroparticles, electromagnetic fields,
and the hybrid fluid. A very simple setup with a laser pulse being reflected from an overdense surface is used for this benchmark. We expect the total energy
\begin{align*}
 E_{tot} =  \sum_l{m_l c^2 (\gamma-1)} &+ \frac{1}{8\pi}\int_V (E^2 + B^2)dV \\
  &+\int_V n_h (\gamma_h -1)m_h c^2 dV
\end{align*}
to be constant, where $m_l$ are the masses of the respective particle species and $\gamma = \sqrt{1 + (p_l/m_lc)^2}$ is the relativistic gamma factor. We denote the hybrid density by $n_h$ and its gamma factor by $\gamma_h$. Figure \ref{energy} shows the total energy of the simulation versus time, which is measured in units of laser periods.
During the laser propagation in vacuum, the energy stays constant except for small fluctuations within the order of magnitude of the machine precision. When the pulse hits the
overdense hybrid plasma surface, it is reflected, as can be seen at the time of 45 laser periods. While this reflection occurs, energy fluctuations are limited by 0.04\% of
the total energy.

\subsection{Target Normal Sheath Acceleration}
\begin{figure}[h]
	\begin{center}
		\includegraphics[width=0.8\columnwidth]{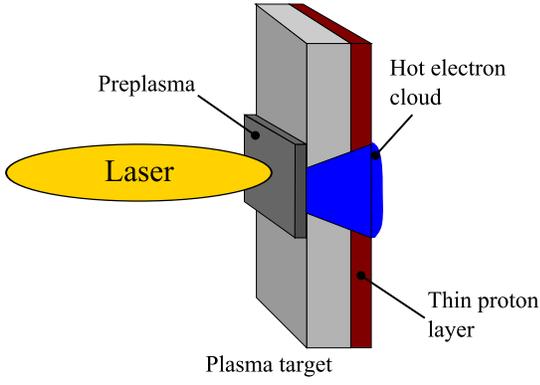}
		\caption{Schematic of the TNSA process.}
		\label{TNSA}
	\end{center}
\end{figure}

For a more realistic benchmark we model a physical setup our hybrid code is very suitable for: We use it for the investigation of the TNSA process. \cite{TNSA}. TNSA provides a possible way of laser ion acceleration out of solids by utilizing the electrostatic fields generated by the space charge of thermal electrons. \\
The process is shown schematically in figure \ref{TNSA}: A $10 fs$
laser pulse of normalized amplitude $a_0 = 2$ is focused on a thin
foil which can be assumed to have been pre-ionized by the laser. The
foil consists of a bulk part of $1000 n_c$, a preplasma on its front
surface, and an 80 nm thick proton layer on its back surface. The
preplasma is modeled as a density ramp reaching from $0$ to $2 n_c$
over a distance of 2 laser wavelengths ($1.6~\mu m$) and treated entirely by the PIC method. Analogously, we use PIC macroparticles for the back surface protons. On the contrary, any attempt to describe the highly overdense main part of the foil as macroparticles would result in numerical problems. Here we use the hydrodynamic feature of H-VLPL, setting the hybrid density on the grid to $1000 n_c$. \\

\begin{figure}[h]
	\begin{minipage}{0.48\columnwidth}
		\begin{center}
			\includegraphics[width=1.0\columnwidth]{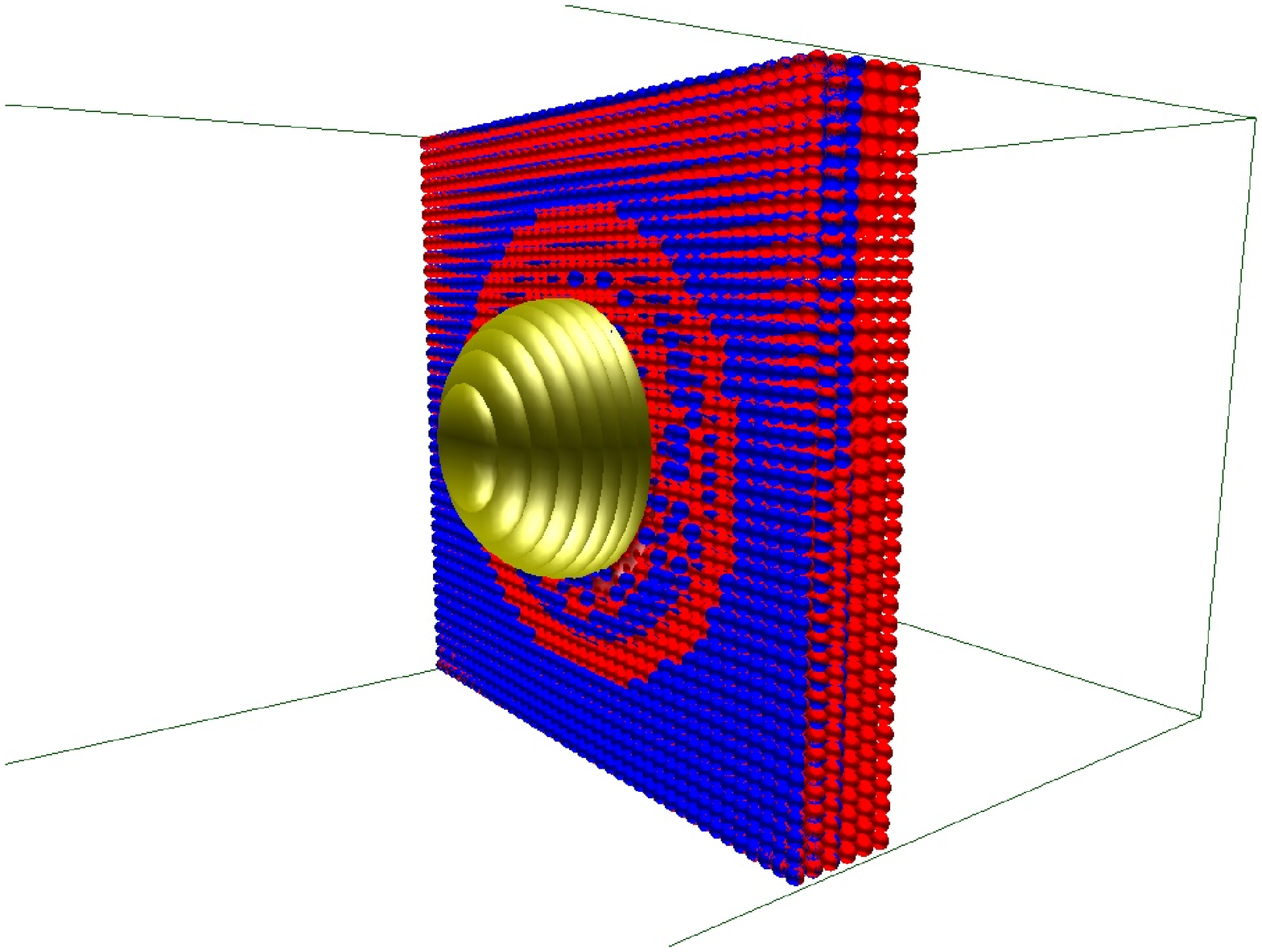}
		\end{center}
	\end{minipage}%
	\begin{minipage}{0.04\columnwidth}
		\hfill
	\end{minipage}%
	\begin{minipage}{0.48\columnwidth}
		\begin{center}
			\includegraphics[width=1.0\columnwidth]{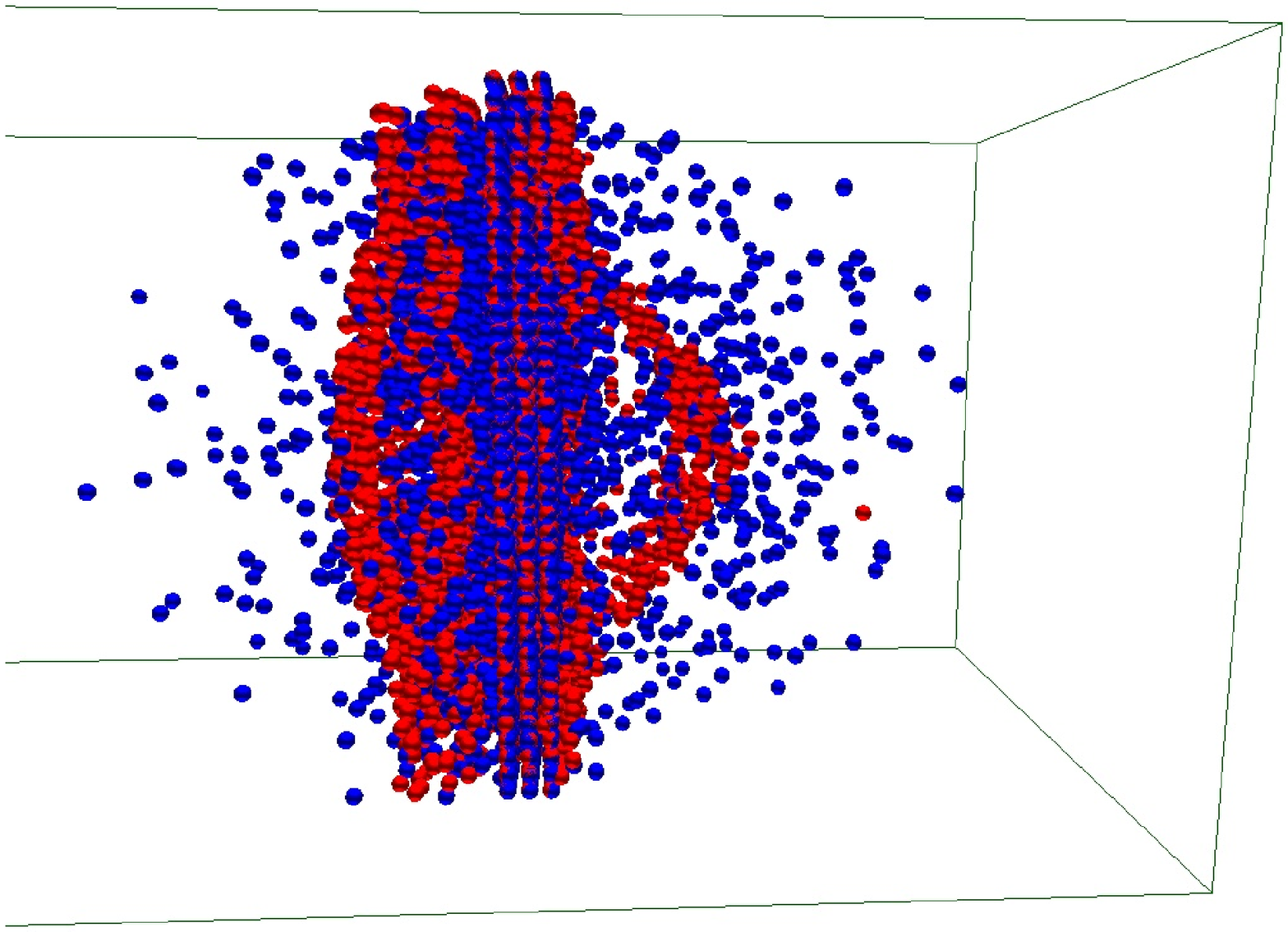}
		\end{center}
	\end{minipage}%
	\caption{Snapshot of the TNSA benchmark simulation after 10 (left) and 380 (right) laser periods. PIC macroparticles containing electrons are displayed blue, while those with protons are rendered red.
		One observes the thin coating of protons dissolving from the back of the foil in the right image.}
\end{figure}

\begin{figure}[H]
	\begin{center}
		\includegraphics[width=0.7\columnwidth]{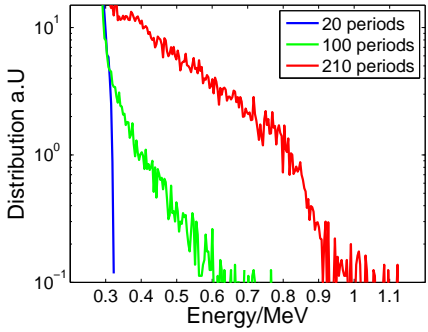}
		\caption{Spectrum of the accelerated ions in the TNSA simulation.}
		\label{spec}
	\end{center}
\end{figure}
The intense laser radiation creates a blow-off region in the front of the foil, resulting in a large cloud of hot electrons, which, in parts, propagates through the foil and passes the coating of the back surface. As the electrons leave the surface, a strong electrostatic field is built up, and the protons are pulled out of the foil and eventually accelerated to high energies. \\
In Fig.~\ref{spec}, the spectrum of the accelerated ions is shown. A maximum energy of about 0.9 MeV is reached, which is quite remarkable considering the laser intensity in the setup. \\
We conclude that our hybrid algorithm succeeded well and efficiently in treating this numerically challenging physical situation. 

\subsection{Comparison of skin depths}
%As a further benchmark for out hybrid code we check whether the ratio of the electric field amplitude at the skin depth $E_s$ to that of
%a reflected wave $E_i$ matches with the linear theory. We set up a sharp edged plasma with densities $n=10$, $100$, and $1000 n_c$. The incident
%wave is a circularly polarized laser pulse with dimensionless amplitude $a_0 = 0.1$ and a duration of 10 fs.
As a further benchmark for our hybrid code we check the decay of a wave in an overdense plasma. According to the linear
theory, it should scale as $E(x) \sim \exp(-x/\delta_s)$ in the plasma, where $\delta_s = c/\sqrt{\omega_p^2 - \omega^2}$ is the skin length.
Several simulations have been set up using a circularly polarized laser pulse with duration $6 \lambda$ and amplitude $a_0 = 0.01$ in order 
to avoid relativistic nonlinearities. The densities of the plasma surfaces used for this benchmark range from $1.5 n_c$ to $500 n_c$.
We show the decay of the wave inside the plasma for three densities; the agreement with the theoretical predictions up to densities of $500 n_c$ is very good. \newline

\begin{figure}[H]
		\begin{center}
			\includegraphics[width=1.0\columnwidth]{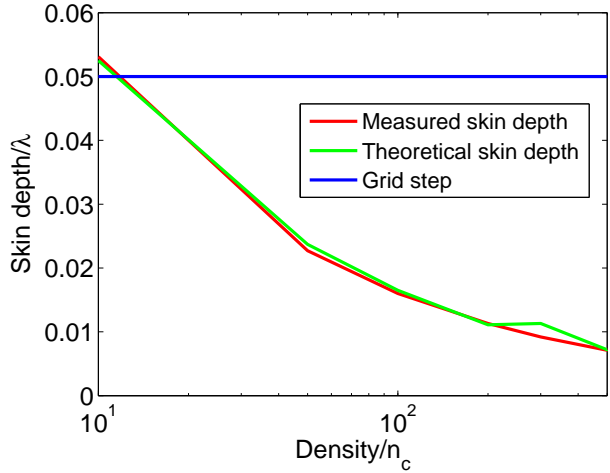}
			\caption{Plot of the skin depth versus the plasma density. The blue line shows the grid step used in the simulation.}
			\label{skindepths}
		\end{center}
\end{figure}

Additionally, by fitting exponentials through the measured field data, one can compute the skin depths of the decay. In figure \ref{skindepths}, the results are shown
and we get an excellent agreement. One has to mention that even though these simulations have been done with a grid step of $0.05 \lambda$,
the skin depths match remarkably well with the theory up to a density of $500 n_c$, where $\delta_s = 0.007 \lambda$.

%\begin{figure}[H]
%	\begin{minipage}{0.48\textwidth}
%		\begin{center}
%			\includegraphics[width=1.0\textwidth]{ExpDecay.eps}
%			\caption{Snapshot of the logarithm of the fields $\log(\sqrt{E_y^2+E_z^2})$ (solid lines) inside the plasma for three different densities. The dashed lines
%				show the theoretical prediction.}
%		\end{center}
%	\end{minipage}%
%	\begin{minipage}{0.04\textwidth}
%		\hfill
%	\end{minipage}%
%	\begin{minipage}{0.48\textwidth}
%		\begin{center}
%			\includegraphics[width=1.1\textwidth]{SkinDepths.eps}
%			\caption{Plot of the skin depth versus the plasma density. The blue line shows the grid step used in the simulation.}
%			\label{skindepths}
%		\end{center}
%	\end{minipage}%
%\end{figure}

\begin{figure}[H]
		\begin{center}
			\includegraphics[width=1.0\columnwidth]{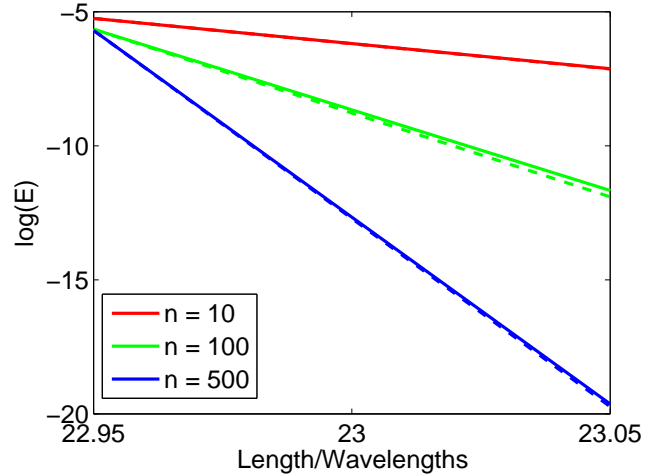}
			\caption{Snapshot of the logarithm of the fields $\log(\sqrt{E_y^2+E_z^2})$ (solid lines) inside the plasma for three different densities. The dashed lines
				show the theoretical prediction.}
		\end{center}
\end{figure}

\subsection{Investigation of Weibel instabilities}
When studying the fast ignition (FI) scenario \cite{FI} in inertial confinement fusion, one is interested in the behaviour of the beam of electrons propagating into the target,
particulary the amount of energy deposited and the shaping of the beam over time. Generally, electron beams running through a background plasma suffer from the major problem of 
the Weibel instability \cite{Wei}, which is a very important issue to be studied if one wants to understand the FI scheme.
The ratio of the beam density to that of the background $n_b/n_p$, as well as the density gradient in propagation direction, is likely to influence the evolution of the beam, its filamentation and the  increase of electromagnetic fields as the instability builds up. \newline
For low densities, roughly about $100 n_c$, PIC simulations can be carried out to perform
these investigations. However, as the electron beam approaches the core of an ICF pellet,
the density will exceed multiple times solid density and conventional PIC codes must be
applied with extremely small time steps in order to avoid numerical instability, and thus
cannot be used with reasonable
computational effort. \newline
We are going to study the phenomenon of the Weibel instability with our new hybrid code, using standard PIC macroparticles for the electron beam and the fluid part in order to model the background plasma. Since H-VLPL has no restrictions for the hybrid densities used, we can perform such simulations within a moderate amount of CPU time. \newline
In order to obtain a proof for the physical correctness of our code within the linear regime, we have launched tests with H-VLPL comparing a classical PIC computation to a hybrid simulation of
this setup. An electron beam with density $n_b$ propagates through a background plasma with $n_p = 100 n_b$. The momentum of the beam electrons is $p_b = mc$ with a thermal spread of $10^{-4} mc$, and the momentum of the background is chosen such that its current compensates for that of the beam plasma, meaning
\begin{align*}
n_b v_b + n_p v_p = 0.
\end{align*}
The setup is restricted to a 2D geometry, with the beams traveling perpendicularly to the $x$-$y$-plane; this is necessary in order to exclude two-stream instabilities.
After about 3.3 beam plasma periods $2\pi / \omega_b$, with $\omega =
\sqrt{4\pi n_b e^2/m}$, one observes a strong filamentation of the
beam, and a magnetic field builds up. When launching the same
simulation with and without the hybrid model, we notice that the
latter succeeds well in describing the filamentation effect at the
initial, linear stage. We compare the integral of the squared magnetic field
\begin{align*}
\int_{V} \textbf{B}^2 dV
\end{align*}
of the two models. At this point it has to be mentioned that during the nonlinear stage of the instability, the present version of H-VLPL will fail in describing the filamentation of the background plasma since it does not treat its continuity equation and convective term of momentum evolution. Additionally, the fluid plasma does not react to magnetic fields directly. \newline 

% \pagebreak

\begin{figure}[H]
\begin{center}
	\begin{minipage}{0.40\columnwidth}
		\begin{center}
			\includegraphics[width=1.0\columnwidth]{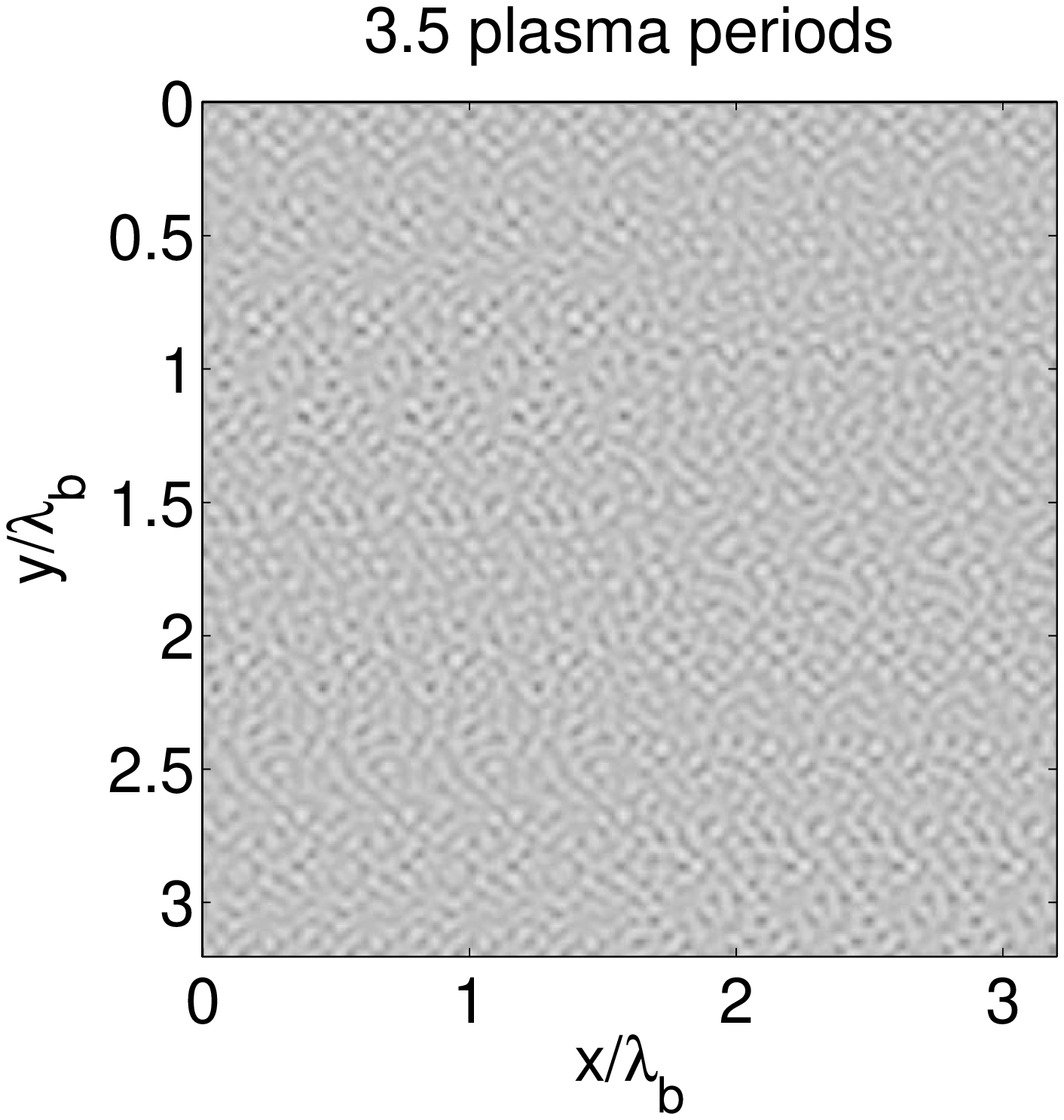}
%			\caption{Snapshot of the logarithm of the fields $\log(\sqrt{E_y^2+E_z^2})$ (solid lines) inside the plasma for three different densities. The dashed lines
%				show the theoretical prediction.}
		\end{center}
	\end{minipage}%
%	\begin{minipage}{0.04\columnwidth}
%		\hfill
%	\end{minipage}%
	\begin{minipage}{0.40\columnwidth}
%		\begin{center}
			\includegraphics[width=1.0\columnwidth]{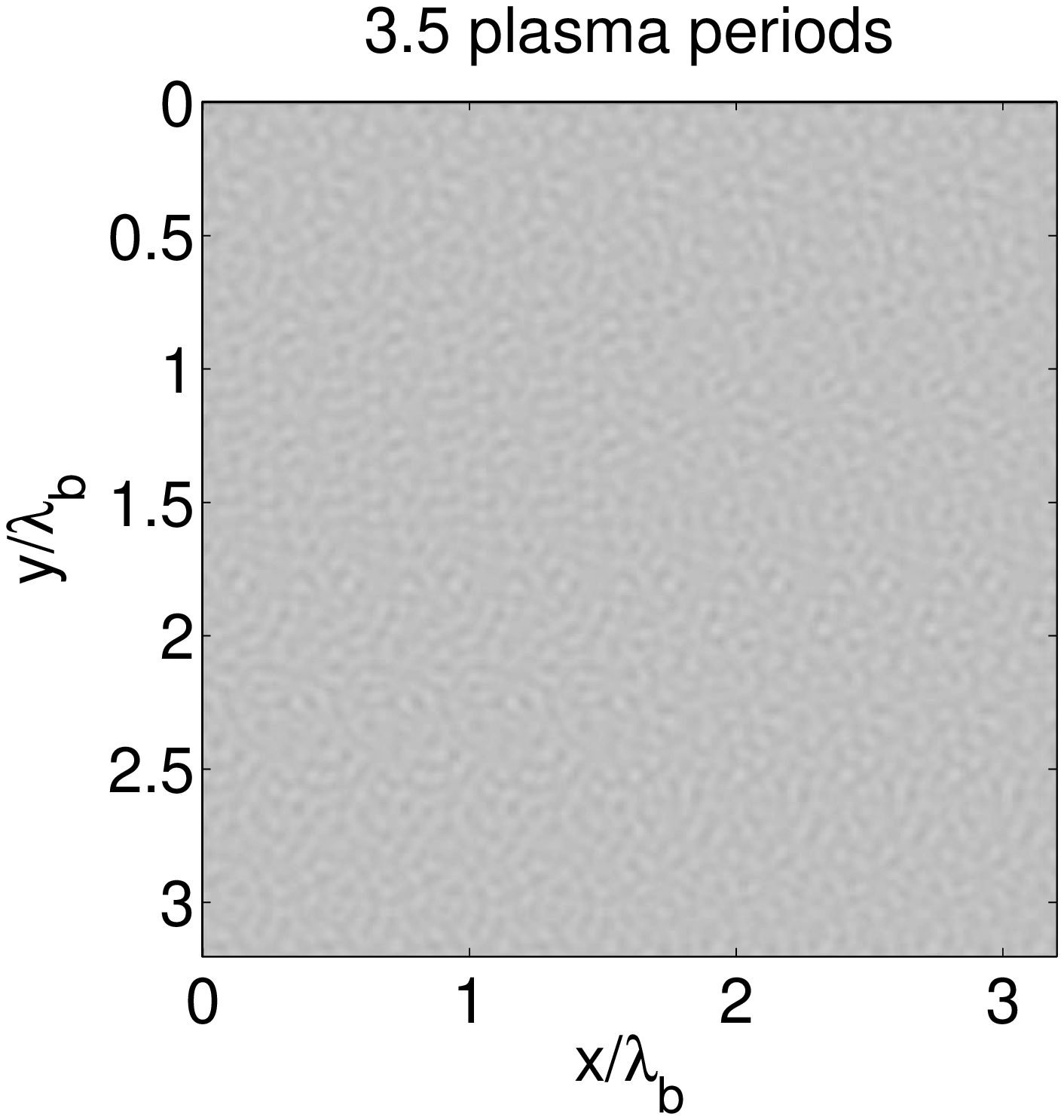}
%			\caption{Plot of the skin depth versus the plasma density. The blue line shows the grid step used in the simulation.}
%		\end{center}
	\end{minipage}%
	\\
%\end{figure}
%\begin{figure}[H]
	\begin{minipage}{0.40\columnwidth}
		\begin{center}
			\includegraphics[width=1.0\columnwidth]{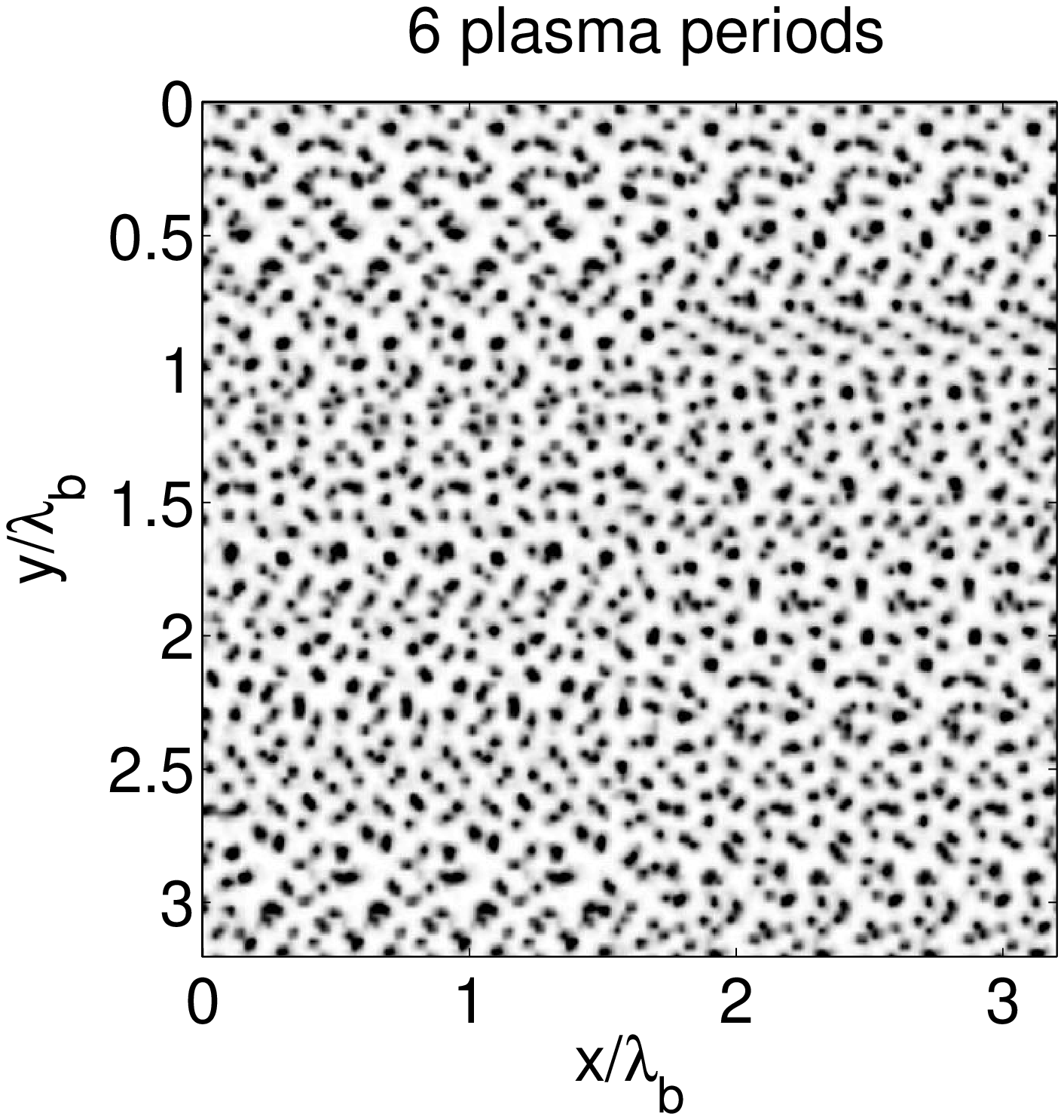}
%			\caption{Snapshot of the logarithm of the fields $\log(\sqrt{E_y^2+E_z^2})$ (solid lines) inside the plasma for three different densities. The dashed lines
%				show the theoretical prediction.}
		\end{center}
	\end{minipage}%
%	\begin{minipage}{0.04\columnwidth}
%		\hfill
%	\end{minipage}%
	\begin{minipage}{0.40\columnwidth}
		\begin{center}
			\includegraphics[width=1.0\columnwidth]{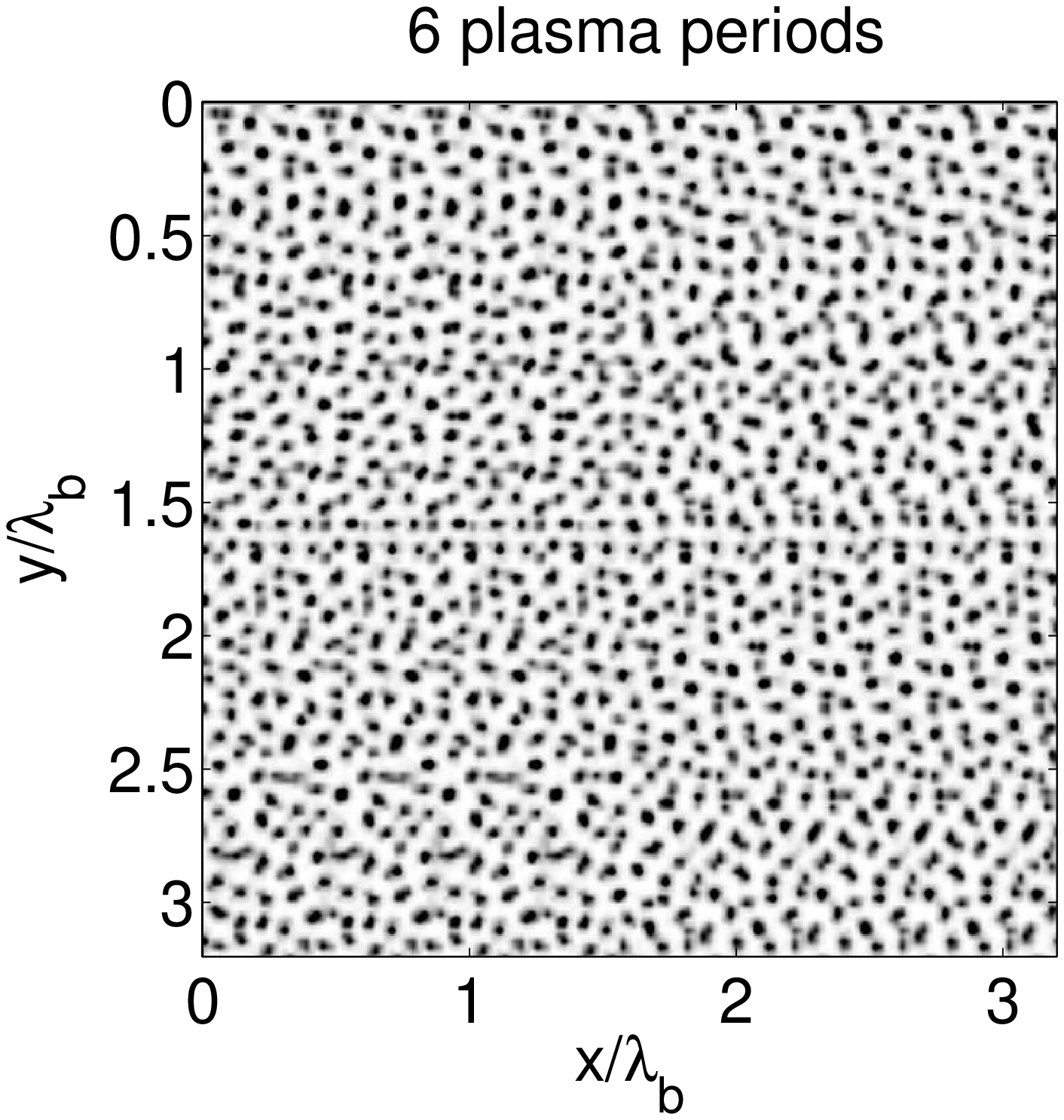}
%			\caption{Plot of the skin depth versus the plasma density. The blue line shows the grid step used in the simulation.}
		\end{center}
	\end{minipage}%
	\\
%\end{figure}
%\begin{figure}[H]
	\begin{minipage}{0.40\columnwidth}
		\begin{center}
			\includegraphics[width=1.0\columnwidth]{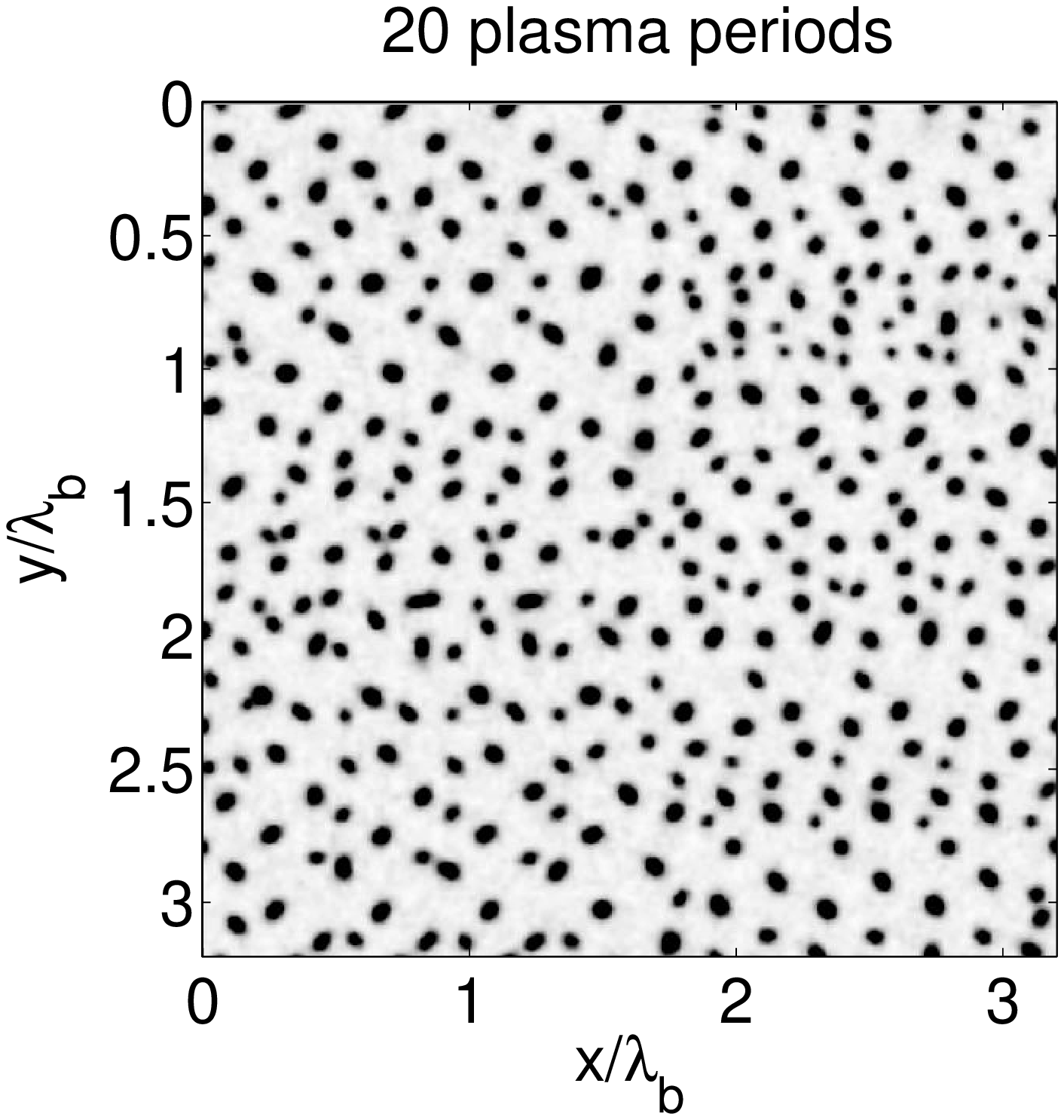}
%			\caption{Snapshot of the logarithm of the fields $\log(\sqrt{E_y^2+E_z^2})$ (solid lines) inside the plasma for three different densities. The dashed lines
%				show the theoretical prediction.}
		\end{center}
	\end{minipage}%
%	\begin{minipage}{0.04\columnwidth}
%		\hfill
%	\end{minipage}%
	\begin{minipage}{0.40\columnwidth}
		\begin{center}
			\includegraphics[width=1.0\columnwidth]{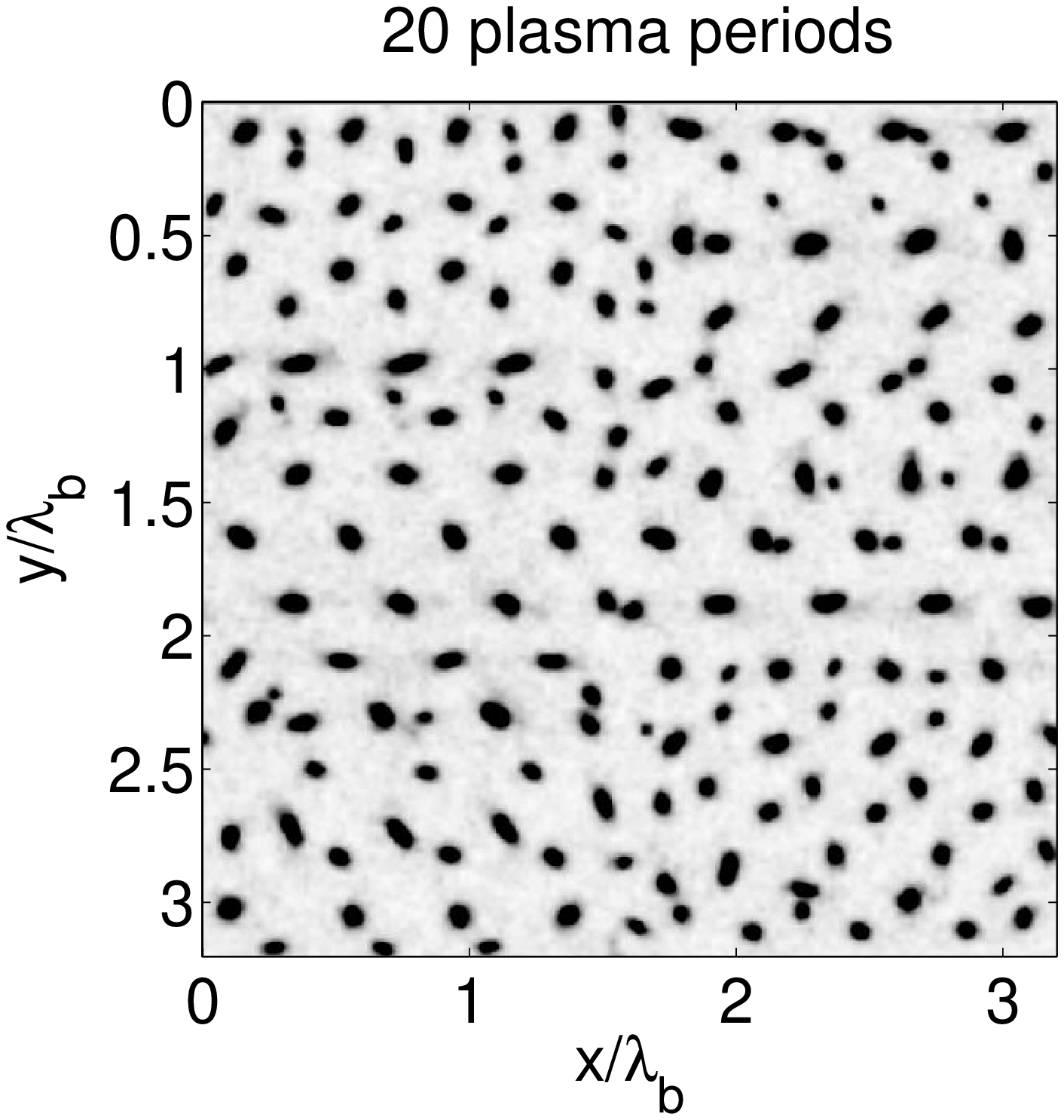}
%			\caption{Plot of the skin depth versus the plasma density. The blue line shows the grid step used in the simulation.}
		\end{center}
	\end{minipage}%
\end{center}
	\caption{Snapshot of the Weibel instability benchmark
	simulations with PIC (left) and the hybrid code (right) at
	3.5, 6, and 20 beam plasma periods. We observe very similar
	behaviour, although the instability starts approximately 0.5
	period later with the hybrid model due to the lower numerical noise.}
\end{figure}

\begin{figure}[h]
	\begin{center}
		\includegraphics[width=1.0\columnwidth]{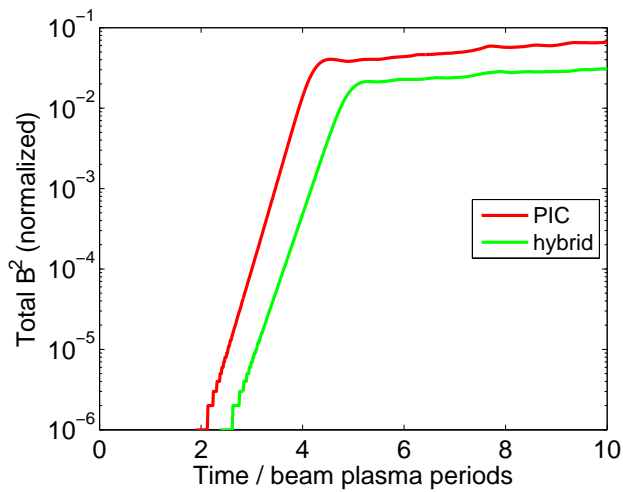}
		\caption{Integral of the squared \textbf{B}-field over the simulation plane. During the linear stage of the instability, we observe an exponential growth with almost the same
			growth rate.}
	\end{center}
\end{figure}

Nevertheless, the behaviour, and the growth rate of the Weibel instability during the linear stage are accurately reproduced. This result indicates the applicability of H-VLPL to the Weibel instability scenario, and makes further investigations of the effect with an advanced, fully hydrodynamic hybrid code appear promising.

\section{Outlook}
The next step in the development of our hybrid laser plasma simulation
system H-VLPL will be the full nonlinear hydrodynamic description of
the background plasma. This includes the continuity equation to
describe the fluid transport as well as momentum transport equation. 
We are going to further study the physical effects mentioned above,
namely the TNSA process and the Weibel instability, using an advanced
version of H-VLPL, which is currently under development. 

\section*{Acknowledgement}
This work has been supported by the Deutsche Forschungsgemeinschaft
via GRK 1203 and SFB TR 18.

{}

\end{document}